\documentclass[apj]{emulateapj}
\usepackage{amssymb}
\usepackage{epsfig}
\usepackage{longtable} 
\usepackage{amsmath}
\usepackage{amsfonts}
\usepackage{subfigure}
\usepackage{verbatim}
\usepackage{hyperref}
\usepackage{filecontents}
\newcommand{\lp}{\left (}
\newcommand{\rp}{\right )}

\newcommand{\mic}{\, \mu \text{m}}
\begin{document}


\title{Finding rocky asteroids around white dwarfs by their periodic thermal emission}

\author{Henry W. Lin\altaffilmark{1}, Abraham Loeb\altaffilmark{2}}
\email{henrylin@college.harvard.edu}
\altaffiltext{1}{Harvard College, Cambridge, MA 02138, USA}
\altaffiltext{2}{Harvard Astronomy Department, Harvard University, Cambridge, MA 02138, USA}

\begin{abstract}
Since white dwarfs are small, the contrast between the thermal emission of an orbiting object and a white dwarf is dramatically enhanced compared to a main sequence host. Furthermore, rocky objects much smaller than the moon have no atmospheres and are tidally locked to the white dwarf. We show that this leads to temperature contrasts between their day and night side of order unity that should lead to temporal variations in infrared flux over an orbital period of $\sim 0.2$ to $\sim 2$ days. Ground based telescopes could detect objects with a mass as small as $1\%$ of the lunar mass $M_L$ around Sirius B with a few hours of exposure. The James Webb Space Telescope (JWST) may be able to detect objects as small as $10^{-3} M_L$ around most nearby white dwarfs. The tightest constraints will typically be placed on 12,000 K white dwarfs, whose Roche zone coincides with the dust sublimation zone. Constraining the abundance of minor planets around white dwarfs as a function of their surface temperatures (and therefore age) provides a novel probe for the physics of planetary formation.

\end{abstract}

\maketitle

\section{Introduction} 
	The search for extrasolar planets around main sequence stars has already harvested thousands of candidates over the past decade \citep[see e.g.][]{kepler13}. Although the number of confirmed exoplanets is now well over a thousand, no exomoons have yet been confirmed, though the smallest exoplanet detected has twice the mass of the moon \citep{konacki03}. To date, no detections of extrasolar rocky objects with masses an order of magnitude lower than a lunar mass $M_L = 7.35 \times 10^{25}$ g have been reported.

	In this Letter, we propose a method to detect objects with masses several orders of magnitude smaller than $M_L$ in close orbits around white dwarfs (WDs). WDs are ideal for two reasons. First, the small size of a WD compared to that of a main sequence star makes the detection of faint companion objects and their properties much easier \citep{burl02, loebmaoz, lin}.
	Second, evidence for circumstellar rocky material already exists. The presence of metal enrichment on the surface of many WDs suggests accretion of rocky debris, since the sedimentation timescale of WDs is often orders of magnitude shorter than their age. The inferred metal accretion rates cannot be accounted for by the interstellar medium \citep{farihi2010}. Furthermore, at least 30 WDs show evidence for dusty disks that are within the Roche zone of the host star \citep{farihi10, koester14, bergfors14}. The ``standard model" (see reviews by \citet{farihi11} and \citet{debes}) for these disks is that they originate from a close-in minor planet that was perturbed by an outer planet into the Roche zone, where it disintegrated into a disk \citep{jura03}. Alternatively, a small fraction of the disks might be accounted for by scrambled comets \citep{comet1, comet2, comet3, stone14}, though the metal abundance of most of the dusty disk WDs favors tidal disruption of rocky objects.

	Consistency with the dusty disk model places lower limits on the population of rocky objects around WDs. \citet{wyatt14} developed a theoretical model where an underlying mass distribution of asteroids gradually accrete onto WDs, and concluded that to explain photospheric absorption lines, asteroids of masses $\gtrsim 5\times 10^{-2} M_L$ are required for typical WDs. The minimum total mass required to explain the metal pollution alone in one DBZ WD is $\sim 0.01 M_L$ \citep{dufour10}, roughly the mass of Ceres, the largest asteroid in the asteroid belt. With recent work estimating that $\gtrsim 30\%$ WDs with an age of 20--200 Myr accrete planetary debris \citep{koester14}, a method to directly detect such objects should either lead to many detections or a revision of the current paradigm.%Finding asteroids $\sim 10^{-3} M_L$ around WDs is on.

%The mass of the dusty disks is estimated to be of order $10^{-4} M_\oplus$ in DAZ WDs $T \sim 10^4 K$. Roughly $10 \%$ of DAZ WDs have dusty disks. %The absence of dusty disks in hot WDs could be attributed to the presence of small rocky bodies orbiting 

	This paper is organized as follows. In \S 2, we study the properties of objects with masses much less than $M_L$ arounds WDs. In \S 3, we calculate the minimum size of objects that could be detected with existing telescopes and the future James Webb Space Telescope (JWST)\footnote{\url{http://www.stsci.edu/jwst/}}. Finally, we discuss possible issues as well as the significance of detection (or upper limits) in \S 4, and summarize our conclusions in \S 5.

%%%%%%%%%%%%%%%%%%%%%%%%%%%%%%
\section{Properties of WD Companions with $M \ll M_L$}
%%%%%%%%%%%%%%%%%%%%%%%%%%%%%%
\begin{comment}
	The distribution of asteroids in protoplanetary disks often follows a power law in mass, $M$,
\begin{equation}
	\frac{d N}{d M} \propto M^{-q},
\end{equation}
	where $q = 11/6$ if the asteroids formed by colliding with a dispersal threshold independent of mass \citep{dohn69}. \citet{wyatt14} infer $q \sim 1.6$ for metal-rich WDs, implying that most objects are low mass.%For In the solar system, there are roughly $\sim 10$ objects in the logarithmic band centered around the mass of the moon $M_L$. Therefore, we expect roughly $10^4$ objects with mass $>10^{-3} M_{L}$.
\end{comment}

	The tidal locking timescale for a gravitationally bound rocky object with $M < M_\oplus$ at an orbital radii $< 0.02$ AU around a WD is $\lesssim 1000$ yr \citep{agol11},
	\begin{comment}
	\citep{murray}
	\begin{equation}
	t_\text{lock} \sim 30 \, \text{yr} \lp \frac{a}{0.01 \text{AU}}\rp^6 \lp \frac{M}{10^{-3} M_L}\rp^{2/3},
	\end{equation}
	\end{comment}
	i.e., much shorter than the planet formation timescale of $\sim 1$ -- 10 Myr for the objects under consideration. We thus expect that virtually all asteroids should be tidally locked to the host WD.

	Over the age of the solar system, only objects with a mass $\gtrsim 0.1 M_L$ can retain volatiles at 50 K \citep{schal07}. At higher temperatures, a much larger mass is needed, since the timescale for the atmosphere to leak away is dominated by an exponential in inverse temperature, $t_\text{leak} \sim \exp (GM\mu/kTR)$, where $\mu$ is the atomic weight and $R$ is the radius of the object\footnote{The requirement for an object to be spherical is that self-gravity of the object should overwhelm its cohesive strength $\tau_0$ which has units of pressure: $GM^2/R^4 \gtrsim \tau_0$. With $M \sim \rho R^3$, we obtain a critical mass for sphericity $M_\circ = \rho^{-2} \lp{\tau_0}/{G}\rp^{3/2}$.
	For a rocky body with a composition similar to that of the moon, $\tau_0 \sim 10^4$ kPa and $\rho \sim 3\,\text{g\,cm}^{-3}$. Thus we expect objects with $M \gg 10^{-4} M_L$ to be spherical. A more detailed analysis \citep{sphere95, potato} refines $M_\circ$ by less than an order of magnitude.}. Therefore, the objects under consideration here are unlikely to have atmospheres.

	In the absence of an atmosphere, the temperature differences between the night and day sides of an object must be of order unity, because thermal conduction is inefficient. The situation is more dramatic than in the case of our moon, because one side of the object will be tidally locked with the WD illumination. The power transmitted by thermal conduction is

\begin{equation}
	P_\text{c} \sim Q R^2 \sim \kappa \nabla T R^2 \sim \kappa T R,
\end{equation}
	whereas the power radiated due to blackbody radiation is of order
\begin{equation}
	P_\text{r} \sim \sigma T^4 R^2.
\end{equation}
	For an object at 600 K with mass $10^{-3} M_L$ and a thermal conductivity $\kappa \sim 4 \times 10^{3} \,\text{erg\,s}^{-1}\,\text{cm}^{-1}\text{K}^{-1}$, radiation is 7 orders of magnitude more efficient than conduction.
	
	Another way to decrease the contrast between the day and night side is by tidal heating of the objects. The power $P_\text{tides}$ generated by the oscillating tidal forces as the object cycles from apoapsis to periapsis is \citep{barnes09}
	
	\begin{equation} P_\text{tides} = \frac{63}{16 \pi}\frac{(G M)^{3/2} M R^5}{Q'_p a^{15/2}}e^2,
	\end{equation}
	where $Q'_p \sim 10^2$ is the minor planet's tidal dissipation function \citep{q66}. For $e < 1$ and a semi-major axis $a \sim 0.01$ AU, $P_\text{tides}$ is 15 orders of magnitude less than $P_c$. %For $e \lesssim 1$, we find that this defines a habitable zone out to $a \lesssim 0.2$ AU, a factor of 10 increase in the size of the habitable zone from \citet{agol11}.

	The upper bound on the temperature of the object is obtained at the Roche radius $r_R \sim 0.005 \, \text{AU} \lp M_\text{wd}/ 0.6 M_\odot \rp^{1/3}$, giving
	\begin{equation}
		T_r \equiv T_\text{wd} \lp \frac{r_R}{R_\text{wd}}\rp^{-1/2} \sim 0.1 \, T_\text{wd} \lp \frac{M_\text{wd}}{ 0.6 M_\odot} \rp^{-1/6},
	\end{equation}
	where $T_\text{wd}$, $R_\text{wd}$, and $M_\text{wd}$ are the temperature, radius, and mass of the WD.
	
	Dust and small rocky objects are expected to disintegrate at the rock sublimation temperature of $\sim 1200$ -- 1500 K \citep{kob11}. Motivated by dusty debris observations, we target asteroids which do not contain a significant amount of ice, since these will sublimate at even lower temperatures.

\section{Observational prospects}

\begin{figure}
	\centering
	\includegraphics[scale=0.5, clip=true]{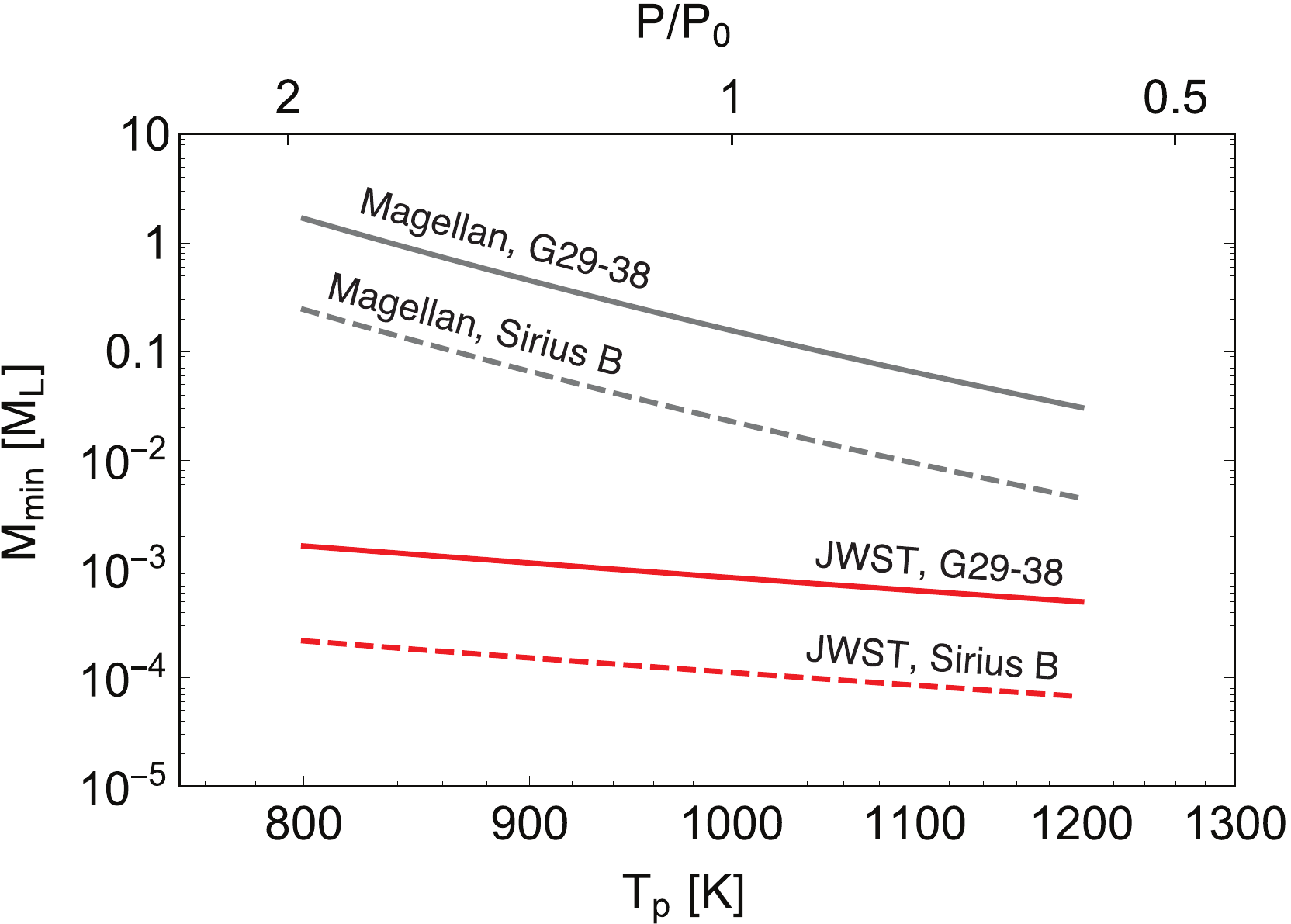}%Left Bottom Right Top
	\caption{Minimum asteroid mass $M_\text{min}$ that Magellan FIRE or JWST MIRI could detect for both Sirius B and G29-38 with a $\sim 1$ hr exposure as a function of the minor planet temperature $T_p$. On the upper axis, we plot the normalized orbital period $P/P_0$ of the asteroid. For Sirius B, $P_0 \sim 70\text{\,hr}$ and for G29-38, $P_0 \sim 7 \text{\,hr}$. Ground based telescopes like Magellan are only sensitive out to a wavelength of $\sim 2.5 \mic$, and thus quickly lose sensitivity to cooler asteroids. Limited photometric stability of JWST may increase $M_\text{min}$.} 
\end{figure}

	Next, we show that it is feasible to observe objects with a mass as low as $10^{-3} M_L$ with JWST. The temperature $T_p$ of a minor planet with exposed area $A \sim \pi R^2$ separated by a distance $d$ from a WD with temperature $T_\text{wd}$ is given by balancing the heating from the WD illumination with cooling from blackbody radiation: $\sigma T_p^4 A = \sigma T_\text{wd}^4 4 \pi R_\text{wd}^2 A/(4 \pi d^2)$, where we neglect the albedo $\sim 0.1$ for a rocky object. This yields $T = T_\text{wd} \lp{R_\text{wd}}/{d}\rp^{1/2}$.
	At the Roche zone orbital separation of $\sim 0.005$ AU, $T_p$ could reach $\sim 0.1 \, T_\text{wd}$. %Since the luminosity of any blackbody scales with $T^4$, any search for thermal emission will be biased towards hotter objects, since our signal will be competing against the Poisson noise from the host WD. Furthermore, 
	For a fixed orbital distance, the signal-to-noise ratio $S/N$ depends on the flux of the WD as follows:

\begin{equation}
	(S/N)^2 \propto T_\text{wd}^4/d_\text{wd}^2,
\end{equation} 
	where $d_\text{wd}$ is the distance from Earth to the WD. The distribution of WD temperatures can be written as $dN/dT_\text{wd} = \dot{N}/\dot{T}_\text{wd}$. Assuming self-similarity in the temperature profile as the WD cools, $\dot{T}_\text{wd} \propto T_\text{wd}^4$. For a constant production rate of WDs, $\dot{N} = \text{const}$, this yields $dN/dT_\text{wd} \propto T_\text{wd}^{-4}$, which is in reasonable agreement with observed WD statistics \citep{loebmaoz}. Since the number of WDs out to a distance $d_\text{wd}$ scales as $N \sim d_\text{wd}^3$, this implies that the minimum distance needed to observe a white dwarf with temperature $T$ scales like $d_\text{wd}^3 \propto T_\text{wd}^{3}$, or

\begin{equation}\label{snt}
	S/N \propto T_\text{wd}^{2}.
\end{equation}
	Therefore it is typically advantageous to target hot WDs, even though they are rarer and therefore farther away. Note that equation (\ref{snt}) applies to asteroids whose temperatures are Roche limited. In the Rayleigh-Jeans part of the spectrum, the flux from the asteroid will oscillate with an amplitude $\sim \Delta T/T \cos i$, where $i$ is the inclination of the orbital plane relative to the line-of-sight. Unless $i \sim \pi/2$, we will see oscillations with a period equal to the orbital period of the object, as the body alternates between exposing its day to night sides. Roughly 100 times more WDs are feasible for this technique compared to transiting systems. Therefore, we can target WDs which are closer by a factor $\sim 100^{1/3}$ than in transit surveys, with a flux higher by $100^{2/3}$, improving the resulting $S/N$ \citep{loebmaoz}.

\subsection{Targeting nearby WDs with ground based telescopes}
	The most abundant WDs are a few Gyr old and have temperatures in the $\sim 5000 - 6000$ K range. However, emission from asteroids would be difficult to observe from the ground for cold WDs, since the sky is bright at IR wavelengths $> 3 \mic$, whereas the peak emission from a rocky asteroid is at $\lambda_\text{max} \sim 12 \mic\,(T_\text{p}/300 \text{\,K})^{-1}$. For instruments sensitive to $\lambda < 3 \mic$, only a few systems are hot enough and close enough to be observed from the ground.

	The brightest WD in the sky is Sirius B, with V magnitude 8.3 and an effective temperature of $\sim 26,000$ K. The estimated age of Sirius B, $\sim 120$ Myr \citep{liebert}, is sufficient to allow for planetesimal formation, a process which occurred in our solar system on a $\sim 1$ -- 10 Myr time scale \citep[see e.g.][]{yin02}. Although Sirius B orbits the main sequence star Sirius A, they are separated by 3 -- 11.5 arcsec (8.2 -- 31.5 AU), making it possible to resolve Sirius B alone. With a $\sim 5\%$ duty cycle over $\gtrsim 50$ hrs (for a few hours of total integration time), it will be possible to detect the variable emission from an object with a mass $\sim 10^{-2} M_L$ with the Folded-port InfraRed Echellette (FIRE) on one of the twin 6.5 meter Magellan telescopes. of these objects requires a photometric stability of $\sim 100$ ppm on a few hours timescale. RMS fluctuations of $\sim 1000$ ppm have already been achieved on one minute timescales (see Figure 6 of \citet{bean13}); fluctuations on 2 hour long timescales should be $\sim 1000/\sqrt{120}$ assuming white temporal noise, giving us the desired sensitivity. However, atmospheric effects such as a time-dependent airmass could lead to systematic errors on hour-long time scales, though calibration against nearby stars may alleviate such issues. Although we simulate observations with FIRE, an imager such as the MMT and Magellan Infrared Spectrograph Detection (MMIRS) would achieve virtually identical limits and allow for real-time photometric calibration against other stars. %Longer exposure times will probably not enable much smaller objects, as systematic noise including flat-fielding errors, instability in the detector sensitivity, and atmospheric fluctuations set a photometric precision to about $\sim 100$ ppm.

	Alternatively, the odds of finding an object may be increased around dusty, metal-rich WDs. The most well studied dusty WD is G29-38 (see e.g. \citet{gansicke06}) which has a temperature $T_\text{wd} = 11700$ K and is 14 pc away. G29-38's Roche zone coincides with its sublimation zone, implying that periodic flux variations can occur over exceedingly short timescales $\sim 4$ hr. In Figure 1, we show the minimum mass $M_\text{min}$ that a $\sim 1$ hr observation with FIRE on one of the Magellan telescopes could detect on both Sirius B and G29-38. As the temperature of the objects drops, the minimum mass increases dramatically since FIRE is only sensitive out to a wavelength of $\sim 2.5 \mic$. For G29-38, the required photometric stability is only $\sim 1000$ ppm, as the white dwarf is cooler than Sirius B and Poisson-limited size of the objects is much larger.

	In calculating detectability thresholds, we assumed that the signal to noise for detecting an asteroid is limited by the Poisson noise of photon counts from the host WD:

\begin{equation}
	(S/N)^2 = \frac{A_p^2}{A_\text{wd}}\int \text{d}\lambda \, \frac{f(\lambda, T_p)^2}{f(\lambda, T_\text{wd})}
\end{equation}
	where $f(\lambda,T) \propto t_\text{int}$ is the number of photons per unit wavelength per unit emitted surface area received by the telescope from a blackbody at temperature $T$ over an integration time $t_\text{int}$. $A_p$ and $A_\text{wd}$ are the effective surface areas of the planet and white dwarfs, respectively. The normalization of $f$ is fixed by the brightness of a WD. Requiring $S/N=5$, we solve for the minimum size of a asteroid that can be detected as a function of WD parameters. For real observations, the required exposure time may be greater by a factor of order unity depending on the spacing of the observations in the time domain. Once evidence is found for a asteroid, follow up observations with a similar integration time but with a cadence optimized for the period of the asteroid would give the best statistics.

	Submillimeter observations with ground-based experiments like the Atacama Large Millimeter/submillimeter Array (ALMA)\footnote{\url{http://almascience.eso.org}} could constrain the population of cooler asteroids, though the signals would be periodic on a longer timescale as the orbital period scales $P \propto T_p^{-3}$. Hotter asteroids could also be targeted, but ALMA would then probe the Rayleigh-Jeans part of the spectrum where it is not possible to constrain the temperature of the asteroid and therefore difficult to distinguish it from atmospheric features on the surface of a WD (see \S 4).  
	%Requiring $S/N=5$ (a $5\sigma$ detection), we can detect objects with $A/A_\text{wd} \sim 10^{-3}$ in a one hour exposure. Longer exposures are possible by stacking multiple one hour exposures over several orbits.
\begin{figure}
	\centering
	\includegraphics[scale=0.45, trim = 0cm 0cm 0cm 0cm, clip=true]{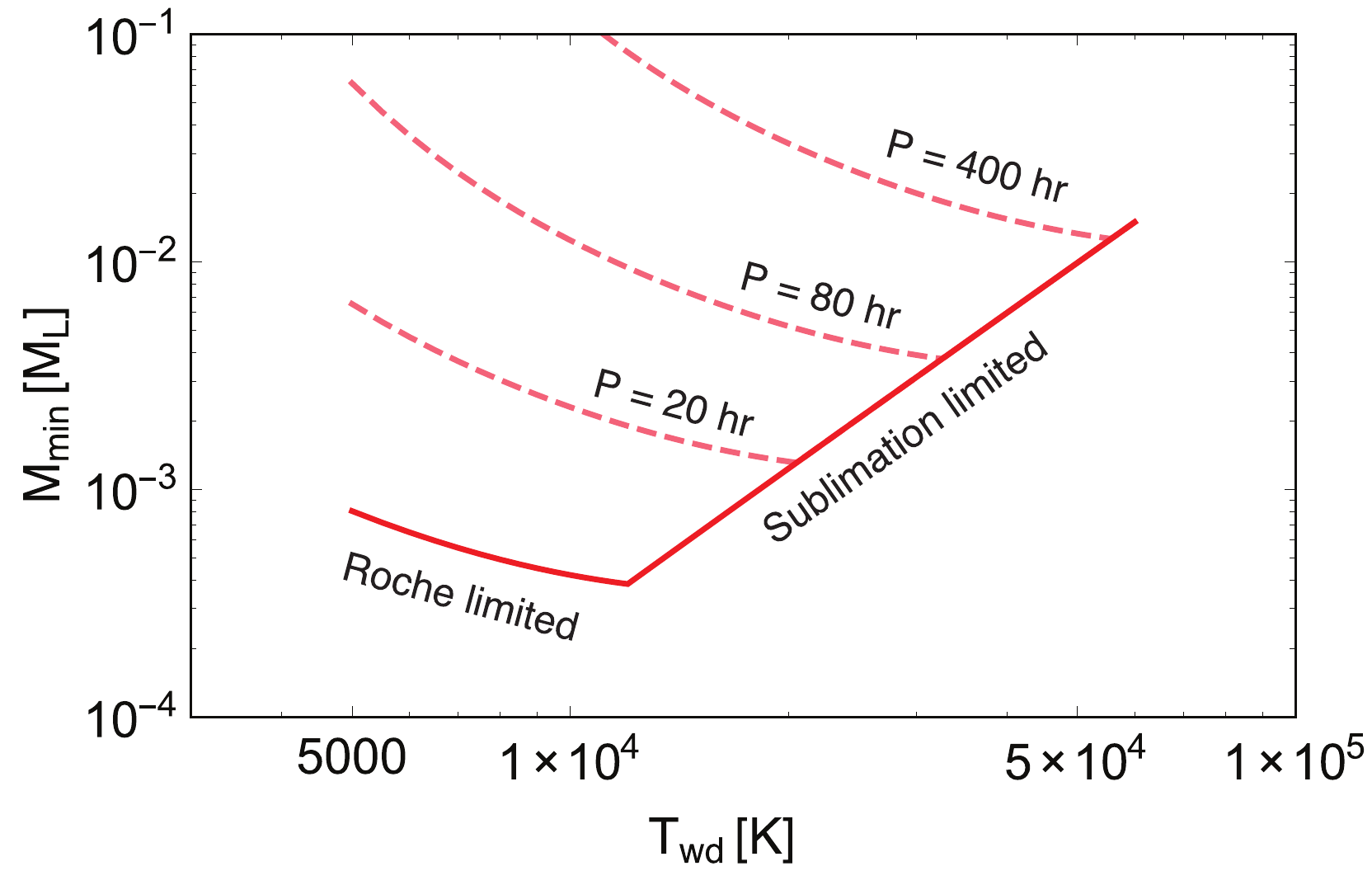}
	\caption{Minimum detectable asteroid mass for nearby WDs at a given temperature $T_\text{wd}$ for a $\sim 1$ hr exposure with JWST MIRI. The solid red curve represents the minimum detectable mass for an object as close to the WD as possible. Dashed pink curves represent constraints on objects with longer orbital periods $P$. At $T_\text{wd} = 1200$ K there is a transition from a Roche limited radius, $r_R$, to a sublimation limited radius, $r_s$. The tightest constraints will typically be placed on $\sim 12,000$ K WDs, where $r_s \sim r_R$.} %from $\sim 4$ hours to $\sim 12$ hours.}
\end{figure}
\subsection{Targeting cool WDs with JWST}
	%We will consider objects at 700 K around a 7000 K WD which are orbiting at roughly $\sim 1$ their Roche radii. By equation \ref{snt}, objects around even hotter WDs will be even easier to detect.
	%An object at 600 K will emit mostly in the mid infrared. 
	A space-based mid-infrared telescope will dramatically improve our constraints on rocky bodies, allowing detection of asteroids around older, cooler WDs. In particular, the Mid Infrared Instrument (MIRI) onboard JWST is sensitive in the 5 -- 28.3 $\mic$ spectral range, which will dramatically decrease the minimum observable mass, especially at lower temperatures where the peak emission is at $\lambda > 2.5 \mic$.

	In Figure 1, we show the minimum mass that a $\sim 1$ hr observation with JWST could detect on both Sirius B and G29-38. Note that the logarithmic slope is much shallower for JWST than Magellan. At $T_p = 800$ K, JWST will improve constraints on asteroid masses by three orders of magnitude if a photometric stability of $\sim 50$ ppm per one-minute exposure is achieved. For cooler, more distant white dwarfs, the required photometric stability is less stringent.

\section{Discussion}
	A potential source of error in future observations might be distinguishing asteroids from atmospheric features on the WD such as spots, although there is currently no evidence \citep{spot12}. One simple way to break the degeneracy is to constrain the effective temperature of the object. In particular, the period of the flux oscillation allows one to predict (up to a factor of order unity due to uncertainty in albedo) the temperature of the asteroid. For a fixed integration time, the fractional uncertainty in the temperature of the minor planet is $\delta T_p/T_p \propto T_p^{-4}$ in the limit where the spectroscopic band is much wider than the wavelength spread in the emitted spectrum of the asteroid. Unless experimental constraints force us to observe only in the Rayleigh-Jeans regime of the spectrum, only a factor of order unity more integration time will be necessary to constrain the asteroid's temperature to a value much cooler than any plausible spot on the WD surface.
	
	If multiple asteroids are orbiting close in, Fourier analysis in the time domain should distinguish the various objects, as asteroids on different orbits will have different periods. Note that the discovery of objects at multiple periods would require be increasingly harder to explain with spots, as a naive model would have all spots rotate with the same period. Furthermore, detection of multiple blackbody components with effective temperatures given by the inferred orbital radii would be a way to confirm the multiple asteroid interpretation.	
	On the other hand, if there are multiple asteroids with very similar periods but different orbital phases, $S/N$  will be reduced as the differing phases wash out the amplitude of the periodic signal. In the limit of a thin dusty disk, the observable periodicity is lost. Consequently, our observational strategy is optimal if asteroids do not spend too long near the Roche zone (such that many asteroids could be in similar orbits) nor too little time near the Roche zone (such that few asteroids will ever be detected). %, as the observed signal is a convolution over the individual periodic signals.
	
	The detection of asteroids around WDs would be a powerful probe of planetary physics. A survey of nearby WDs by JWST could constrain the distribution function $n_p(T_p, T_\text{wd},M_p)$, where $T_p$ and $M_p$ are the temperature and mass of asteroids, respectively. $n_p(T_\text{wd})$ can then be used to infer the statistical evolution of asteroids with time, since $T_\text{wd}$ is linked to the age of the WD by the WD cooling function. %of asteroids as a function of both the asteroid temperature $T_p$ and the temperature of the WD $T_\text{wd}$ could falsify or confirm models of dusty disk formation.
	For example, the tidal disruption model implies that $n_p$ must be intimately tied to the abundance of dusty disks $n_d(T_\text{wd},M_d)$. A generic prediction of this model is that any changes in the mass of debris disks should be accompanied by metal accretion onto the surface of the WD and/or a non-trivial $n_p(T_p)$ dependence. Since constraints have already been placed on $n_d$, observations of $n_p$ could falsify or refine this model.
	
	In addition to detecting minor planets, this technique could also harvest exoplanet detections, though we would expect $\Delta T_p/T_p \ll 1$ for planets with atmospheres. To distinguish a large planet with a small $\Delta T_p/T_p$ from a small object with a larger $\Delta T_p/T_p$, the blackbody spectrum of the object could be obtained and used to infer the effective surface area. Furthermore, the mass of close-in exoplanets could in principle be constrained by observations, as tidal forces should pull the planet into an ellipsoid. As a planet orbits the WD, the effective area of the planet will change over a period $T/2$, as opposed to flux oscillations with a period $T$ from the planet's day/night temperature gradient.
	\\	
\section{Conclusion}
	We have shown that the periodic flux variability from the night/day oscillations of close-in asteroids allows existing and future telescopes to detect objects with a small fraction of the moon's mass. A possible strategy would be to use an existing-ground based telescope to search for asteroids around nearby WDs with $T \gtrsim$ 12,000 K. JWST can then be used to follow up ground-based observations in order to improve the associated constraints by a few orders of magnitude on already-observed targets and to sample the asteroid environments around the entire population of local WDs. The unique environment of WDs thus affords an unprecedented window into the physics of planet formation.

	%One observational strategy to find or constrain the population of objects at $M \gtrsim 10^{-3} M_L$ would be to target the brightest WDs (in apparent magnitude) which show evidence of metal lines or dusty debris disks. Over a period of a day, a few of these objects

\acknowledgements{
We thank Sean Andrews, Jonathan Grindlay, Scott Kenyon, Dani Maoz, and Brian McLeod for useful discussions. This work was supported in part by NSF grant AST- 1312034 and the Harvard College Program for Research in Science and Engineering (PRISE).}
%\begin{comment}

%\end{comment}
%\bibliographystyle{apj}
%\bibliography{bib}{}

\end{document}